\documentclass[%
 reprint,
 amsmath,amssymb,longbibliography
 aps,
 prl,
]{revtex4-1}

\usepackage{graphicx}
\usepackage{amsmath}
\usepackage{rotating}
\usepackage{color}

\newcommand{\sruo}{Sr$_2$RuO$_4$}
\newcommand{\lsoc}{\lambda_{soc}}
\newcommand{\io}{\tilde \mu}

\newcommand{\down}{\downarrow}
\newcommand{\up}{\uparrow}
\newcommand{\spin}{\sigma}

\newcommand{\kv}{{\bf k}}
\newcommand{\qv}{{\bf q}}
\newcommand{\Qv}{{\bf Q}}
\begin{document}

%\title{Superconducting instabilities and Knight shift behavior from spin fluctuations in \sruo}
\title{Fluctuation-driven superconductivity in \sruo ~from weak repulsive interactions}
\author{Astrid T. R\o mer and Brian M. Andersen}
\affiliation{%
Niels Bohr Institute, University of Copenhagen, Vibenshuset, Lyngbyvej 2, DK-2100 Copenhagen,
Denmark
}%

\date{\today}% It is always \today, today,
             %  but any date may be 

\begin{abstract}

We provide results for the leading superconducting instabilities for a model pertaining to \sruo ~obtained within spin-fluctuation mediated superconductivity in the very weak-coupling limit. The theory incorporates spin-orbit coupling (SOC) effects both in the band structure and in the pairing kernel in the form of associated magnetic anisotropies. The leading superconducting phase is found to be $d_{x^2-y^2}$ and a nodal $s$-wave state. However, the odd-parity helical solution can become leading either for small SOC and Hund's coupling  $J$ in the weak $U$-limit, or in the opposite limit with large SOC and $J$ at larger values of the Hubbard-$U$. The odd-parity chiral solution is never found to be leading. Finally we discuss the form of the resulting superconducting spectral gaps in the different explored parameter regimes. 

\end{abstract}

\maketitle

The material \sruo ~has recently attracted significant renewed attention, partly due to a general interest in topological superconducting systems~\cite{Kallin_2016,Mackenzie2017,ImaiSigrist20}, and partly due to the game-changing discovery of a Knight shift suppression upon entering the superconducting state~\cite{Pustogow19,IshidaPreprint19}, contrary to earlier measurements~\cite{Ishida98} and in conflict to the proposal of chiral odd-parity spin-triplet superconductivity. The latter discovery has revived an intense experimental search for determining the nature of the superconducting pairing in \sruo,~and challenged theoretical scenarios for unconventional superconductivity~\cite{Scaffidi2014,Wang19,Gingras18,Zhang18,WangKallin19,RamiresSigrist19}. The fact that \sruo ~in its normal state is a highly anisotropic, relatively weakly correlated Fermi liquid material, gives hope that existing theoretical frameworks for unconventional superconductivity could apply to this material. At present, however, unravelling the ground state pairing structure of \sruo ~constitutes a fascinating open problem in condensed matter physics, a problem that seems to include the complexity of multi-orbital (multi-band) electronic band structure, spin-orbit coupling (SOC), and electron interactions~\cite{Oguchi1995,Haverkort08,Veenstra2014,Tamai19,Mravlje11,Kim2018}.

In a previous publication~\cite{PRL}, we reported the superconducting phases of \sruo ~arising from spin-fluctuation mediated pairing in a framework where SOC is included both in the electronic structure and the pairing interaction. 
We used a realistic SOC, which correctly reproduces the magnetic anisotropy found by neutron scattering and a sizeable Hund's coupling strength\cite{Kim2018}. The pairing interaction was modelled by certain classes of diagrams giving rise to a pairing mechanism of the form of the generalized multi-orbital susceptibility within the random phase approximation (RPA). The result of this procedure pointed towards leading even-parity superconducting phases of either $d_{x^2-y^2}$ or nodal $s$-wave, but large SOC as well as Hund's couplings could prefer an odd-parity helical solution in some cases~\cite{PRL}.

This result is in apparent contradiction to other recent reports of spin-fluctuation mediated superconductivity in \sruo ~reported by Refs.~\onlinecite{Zhang18,WangKallin19} in which the calculations were performed in the weak-coupling regime and in which a helical solution was found at small Hund's couplings $J$. In addition, in Ref.~\onlinecite{WangKallin19} Wang {\it et al.} pointed out  the importance of hybridization between the $xz$ and $yz$ orbitals in favoring helical solutions over chiral pairing.
Motivated by these results, we investigate the role of the Hubbard-$U$ and hybridization, focusing on the
 weak-coupling regime of very small $U$ and $J$. We find that, in agreement with 
Ref.~\onlinecite{WangKallin19}, there is a leading helical solution appearing for small SOC and Hund's couplings in the very weak-coupling regime. However, unlike the reports in Ref.~\cite{Zhang18}, we do not find any leading chiral solutions at any finite Hund's couplings. A finite hybridization between the $xz$ and $yz$ orbitals  produces only small quantitative changes to the phase diagram boundaries compared to the results for the case of zero hybridization. We end the discussion by addressing the magnetic anisotropy and the superconducting spectral gaps in the different superconducting states.

\begin{figure*}[t]
 \centering
   	\includegraphics[angle=0,height=0.19\linewidth]{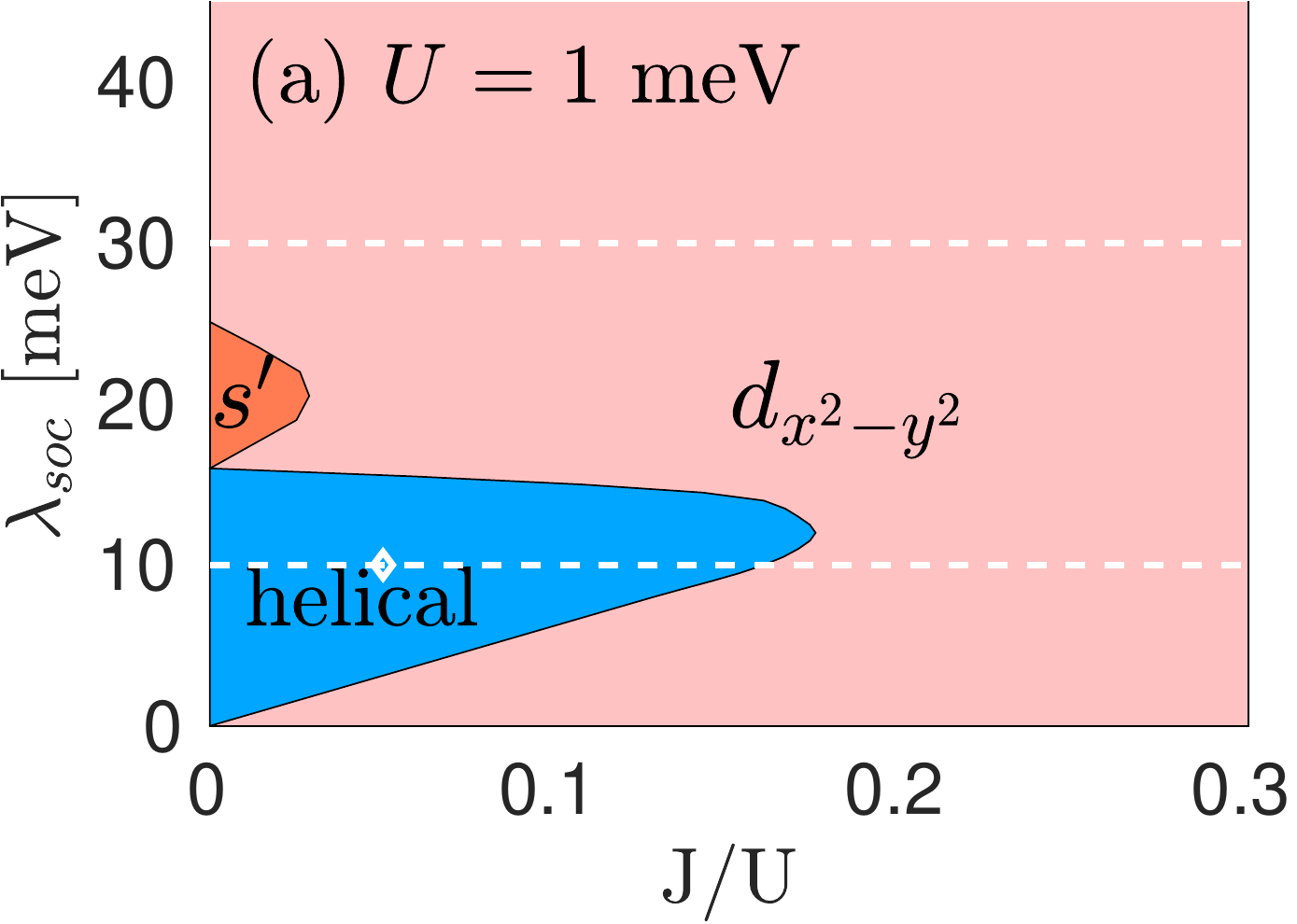}
   	\includegraphics[angle=0,height=0.19\linewidth]{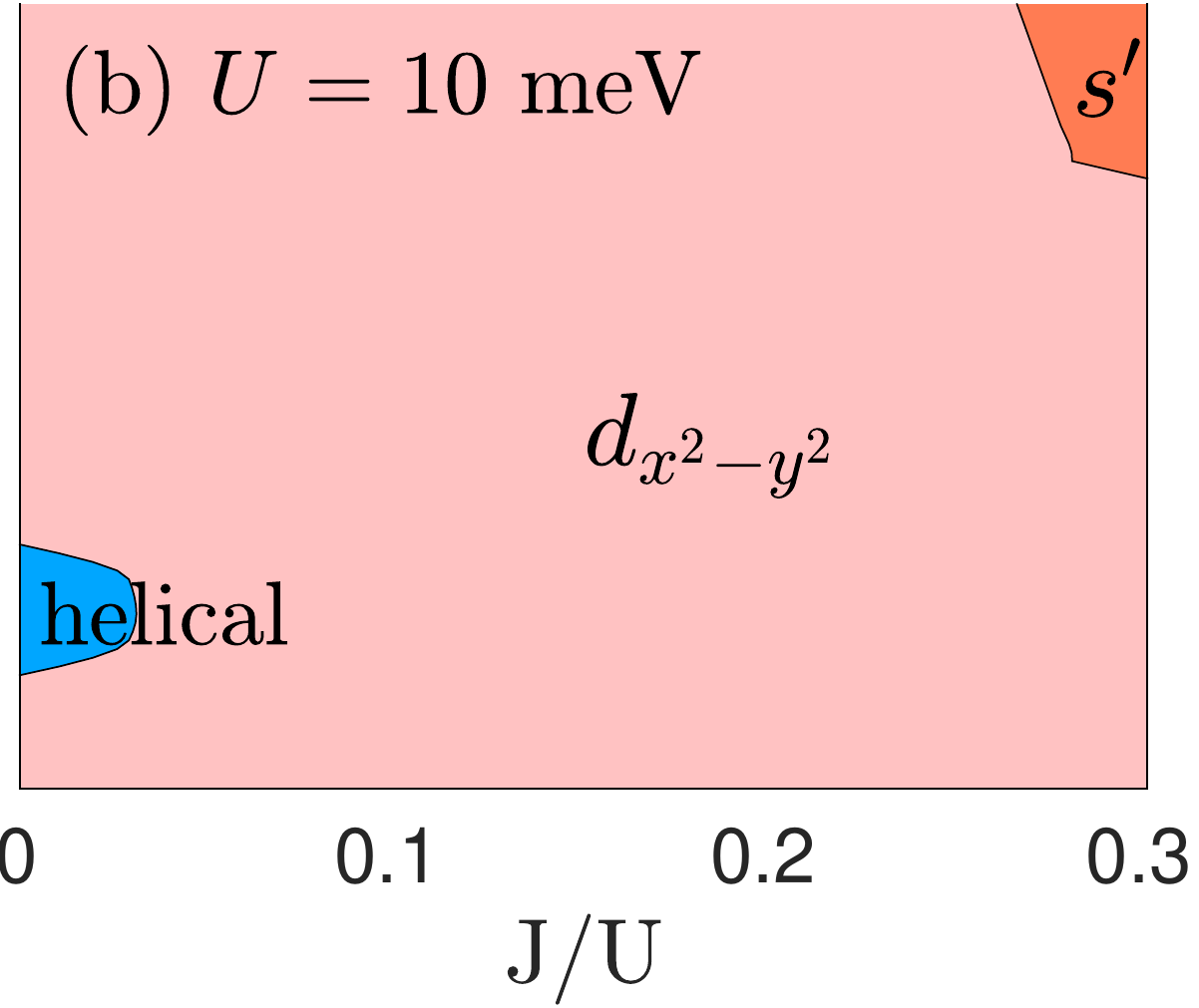}
   	\includegraphics[angle=0,height=0.19\linewidth]{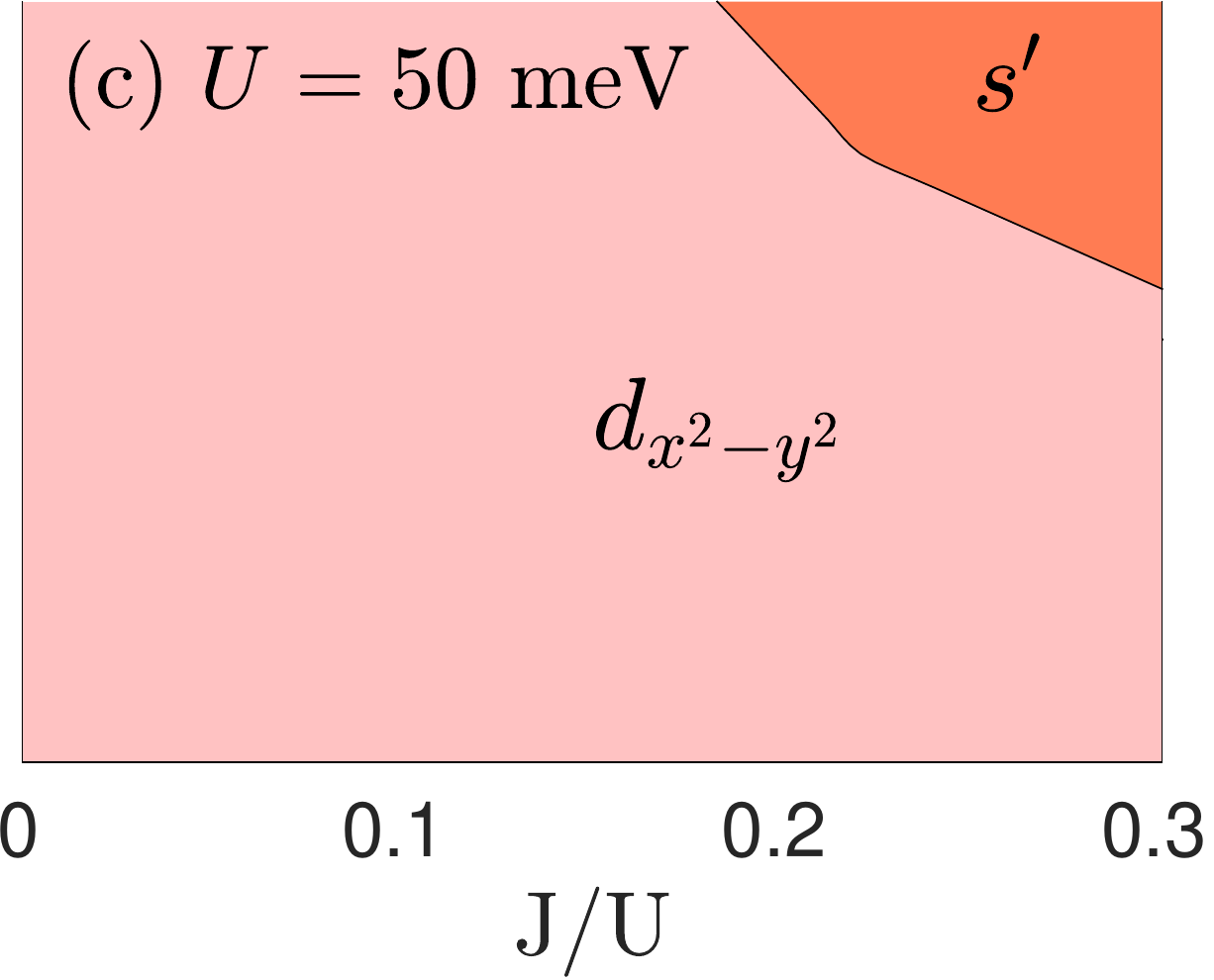}
   	   	\includegraphics[angle=0,height=0.19\linewidth]{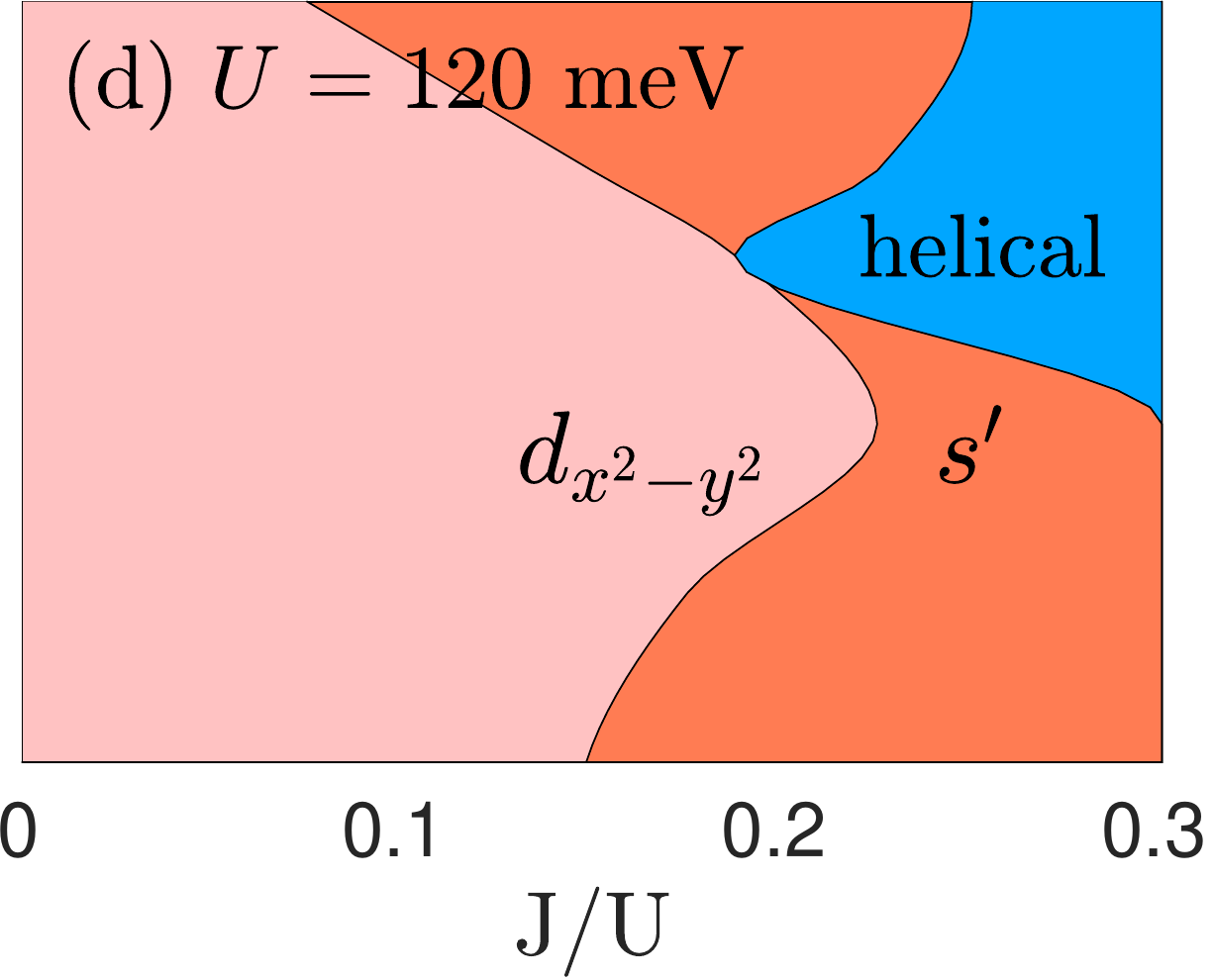}
   	\includegraphics[angle=0,width=0.23\linewidth]{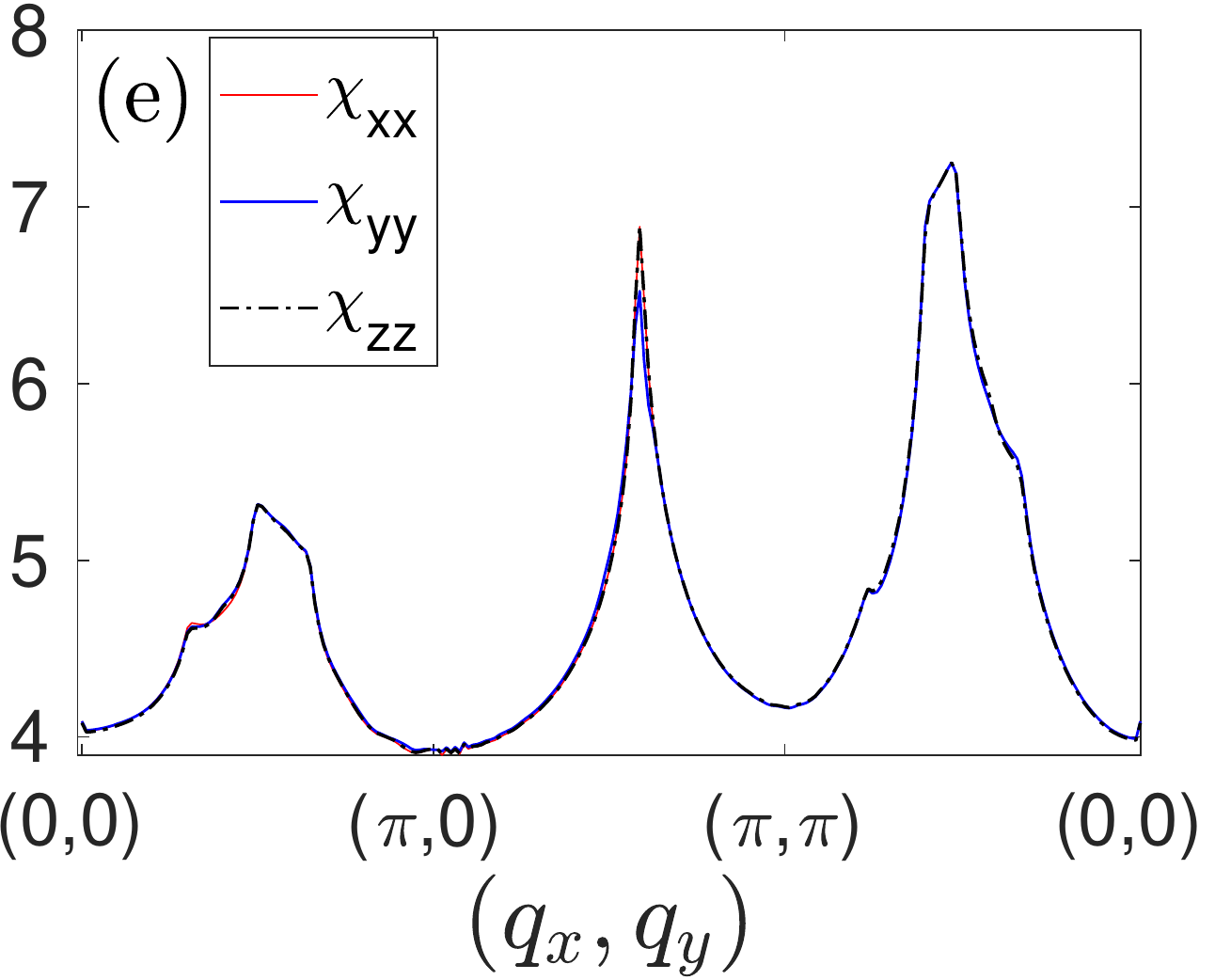}
   	\includegraphics[angle=0,width=0.23\linewidth]{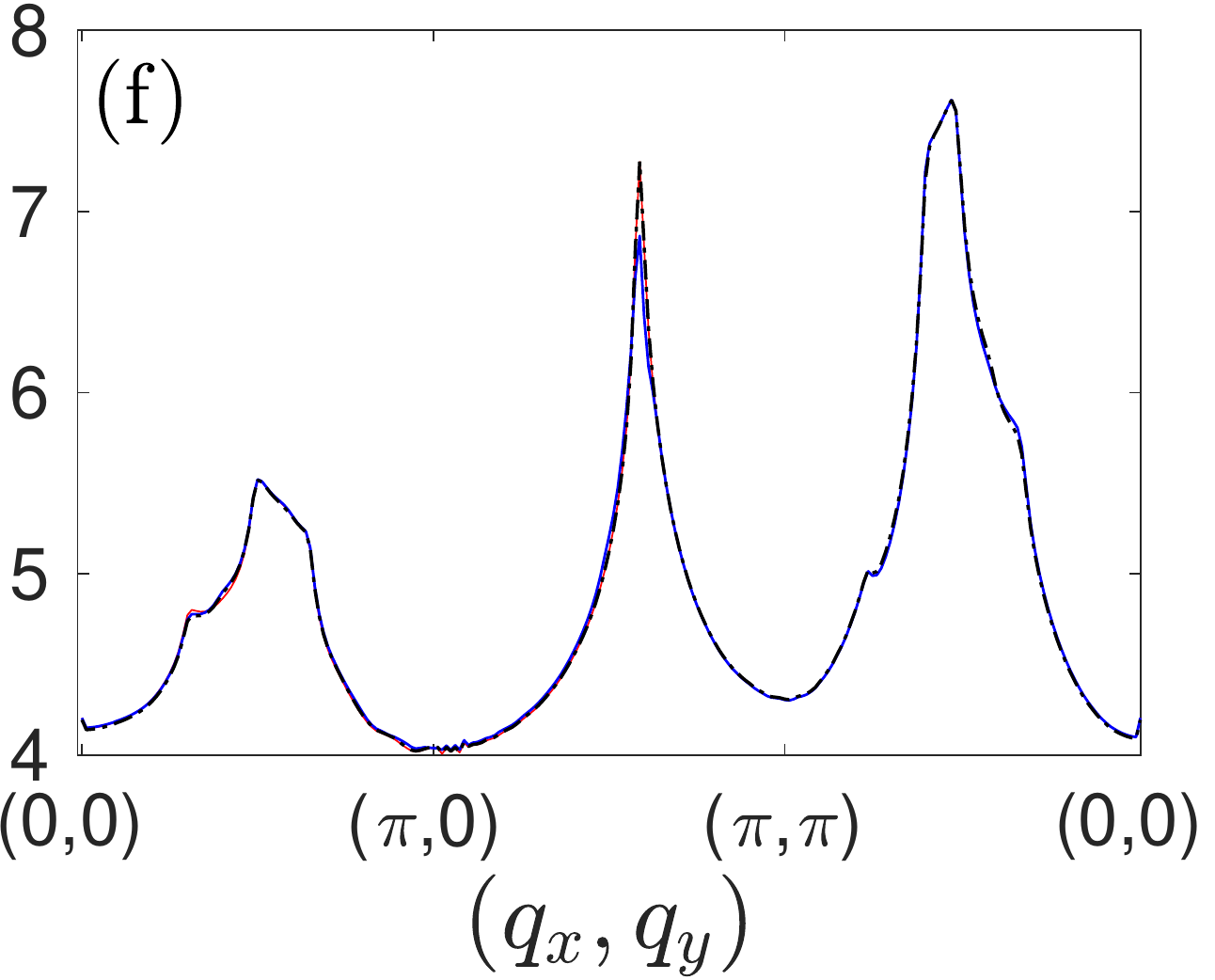}
   	\includegraphics[angle=0,width=0.23\linewidth]{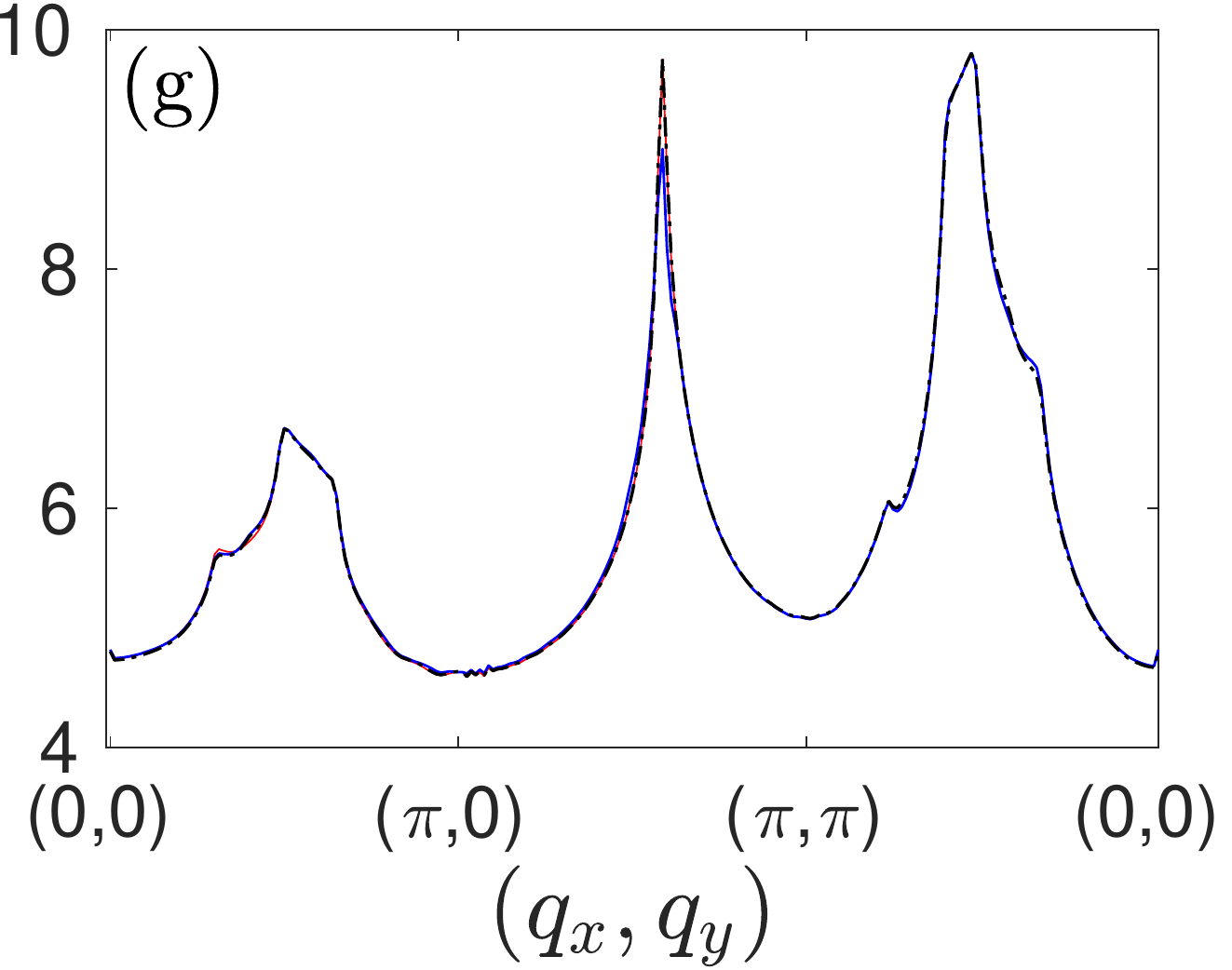}
   	\includegraphics[angle=0,width=0.23\linewidth]{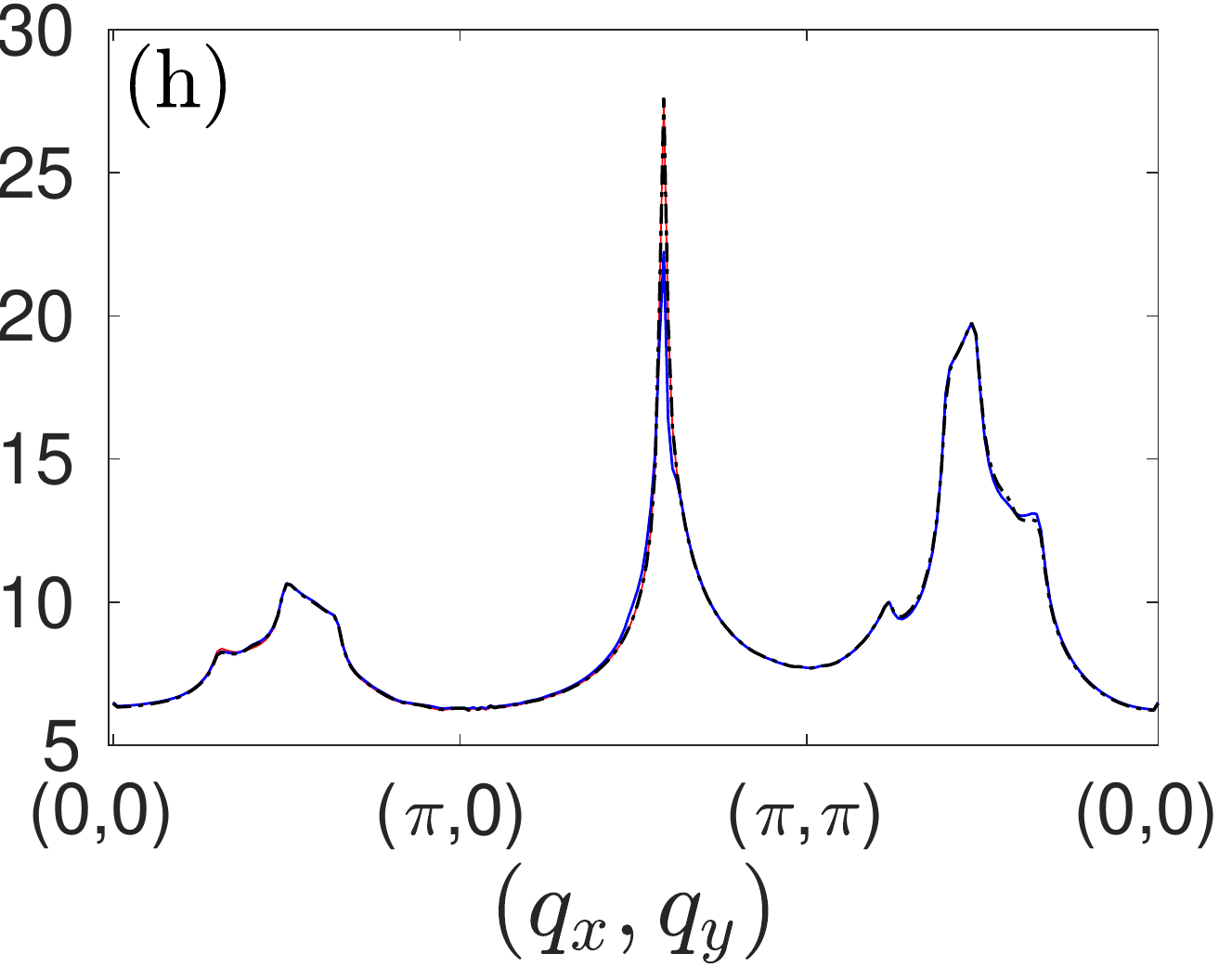}
\caption{(a-d) Leading superconducting instability as a function of spin-orbit coupling $\lsoc$ and Hund's coupling $J$ for increasing value of $U$: $U=1,10,50,120$ meV. The helical solution at small spin-orbit and Hund's coupling is suppressed upon increasing $U$. For larger $U$, a regime of helical solutions appears at large spin-orbit and Hund's coupling, as seen in (d). (e-h) The physical spin susceptibilities, $\chi_\textrm{xx,yy,zz}(\qv,0)$ at $\lsoc=10$ meV and $J/U=0.05$ for increasing Hubbard-$U=1,10,50,102$ meV, respectively. The position in $(J/U,\lsoc)$ is indicated by the white diamond in (a).
The two dashed white lines in (a) display the cuts for which subleading eigenvalues are shown in Fig.~\ref{fig:subgaps}. }
\label{fig:LGE_Cobo}
 \end{figure*} 

The non-interacting Hamiltonian can be written in block-diagonal form $\hat{H}=\sum_\sigma \Psi^\dagger(\kv,\sigma) (H_0+H_{SOC}) \Psi(\kv,\sigma)$ with the matrices $H_0$ and $H_{SOC}$ given by
\begin{eqnarray}
 H_0&=&\left( \begin{array}{ccc}
  \xi_{xz}(\kv) & g(\kv) &0 \\
  g(\kv) & \xi_{yz}(\kv) & 0 \\
  0 & 0 &   \xi_{xy}(\kv)
 \end{array}\right),
\end{eqnarray}
\begin{eqnarray}
 H_{SOC}&=&\frac{1}{2}\left( \begin{array}{ccc}
  0 & -i\spin\lsoc & i\lsoc \\
  i\spin\lsoc & 0 & -\spin\lsoc \\
  -i\lsoc & -\spin\lsoc &  0\\
 \end{array}\right),
 \label{eq:H0Hsoc}
\end{eqnarray}
within the basis
$\Psi(\kv,+)=[c_{xz\up}(\kv),c_{yz,\up}(\kv),c_{xy,\down}(\kv)]$, and 
$\Psi(\kv,-)=[c_{xz,\down}(\kv),c_{yz,\down}(\kv),c_{xy,\up}(\kv)]$.
Here, $c_{\mu,s}(\kv)/c^\dagger_{\mu,s}(\kv)$ are electronic annihilation/creation operators of orbital character $\mu$ and spin $s$. The pseudospin is $\spin=+(-)$ for the up (down) block Hamiltonian stated in Eq.~(\ref{eq:H0Hsoc}).
The electronic dispersions are given by
\begin{eqnarray}
 \xi_{xz}(\kv)&=&-2t_1\cos k_x -2t_2\cos k_y -\mu, \\
 \xi_{yz}(\kv)&=&-2t_2\cos k_x -2t_1\cos k_y -\mu,\\
 \xi_{xy}(\kv)&=&-2t_3(\cos k_x +\cos k_y) \nonumber \\
&&-4t_4\cos k_x \cos k_y-2t_5(\cos 2k_x +\cos 2k_y) -\mu, \nonumber \\
&&
\end{eqnarray}
 and orbital hybridization between $xz$ and $yz$ is parametrized by $t'$ in  $g(\kv)=-4t'\sin(k_x)\sin(k_y)$.
The hopping constants are given by $\{t_1,t_2,t_3,t_4,t_5\}=\{88,9,80,40,5\}$ meV\cite{Cobo16,Zabolotnyy13}, and we set hybridizations to $t'=0$ or $4.4$ meV($=0.05t_1$)\cite{WangKallin19}. The chemical potential is $\mu=109$ meV and the model is restricted to two dimensions.
The effective electron-electron interaction in the Cooper channel from the multi-orbital Hubbard Hamiltonian due to spin fluctuations was derived in Ref.~\cite{PRL}. It includes intra- and interorbital Coulomb interactions and Hund's coupling terms and effective interactions mediated by spin-fluctuations in the multiorbital random-phase approximation:
\begin{equation}
 \hat{H}_{int}=\frac{1}{2}\!\sum_{ \kv,\kv' \{\tilde \mu\}}\!\!\Big[V(\kv,\kv')\Big]^{\io_1 , \io_2 }_{\io_3,\io_4 }  c_{\kv \io_1 }^\dagger  c_{-\kv \io_3 }^\dagger c_{-\kv' \io_2 } c_{\kv' \io_4 },
 \label{eq:Heff}
\end{equation}
with the pairing interaction given by
\begin{eqnarray}
\Big[V(\kv,\kv')\Big]^{\io_1 , \io_2 }_{\io_3,\io_4 } &=&\Big[U\Big]^{\io_1 , \io_2 }_{\io_3,\io_4 }+\Big[U\frac{1}{1-\chi_0U}\chi_0U\Big]^{\io_1 \io_2}_{\io_3 \io_4}(\kv+\kv') \nonumber \\
&& -\Big[U\frac{1}{1-\chi_0U}\chi_0U  \Big]^{\io_1\io_4}_{\io_3 \io_2}(\kv-\kv') .
\label{eq:Veff}
\end{eqnarray}
The label $\io \:= (\mu,s)$ is a joint index for orbital and electronic spin. The real part of the generalized multi-orbital spin susceptibility $\chi_0=[\chi_0]^{\tilde \mu_1,\tilde \mu_2}_{\io_3,\io_4}(\qv,i\omega_n)$ is evaluated at zero energy and includes the effects of SOC. The interaction Hamiltonian is projected to band and pseudo-spin space
\begin{eqnarray}
 \hat{H}_{int} \!=\!\!\!\!\!\!\!\sum_{n,n', \kv,\kv'}\!
 \sum_{l,l'} 
\overline\Psi_{l}(n,\kv)
 ~\!\frac{1}{2}\Gamma_{l,l'}(n,\kv;n',\kv')~\Psi_{l'}(n',\kv'),\nonumber \\
\end{eqnarray}
with $n,n'$ are band indices. The pseudo-spin information is carried by the $l,l'$ indices 
with the fermion bilinear operator
\begin{eqnarray}
 \overline\Psi_l(n,\kv)&=&s_l\beta^\dagger_{\kv n\spin_1}[\Gamma_l]_{\spin_1\spin_2}\beta^\dagger_{-\kv n'\spin_2}\delta_{n,n'}, \nonumber \\
\Psi_l(n,\kv)&=&\beta_{\kv n\spin_1}[\Gamma_l]_{\spin_1\spin_2}\beta_{-\kv n'\spin_2} \delta_{n,n'},
\label{eq:Psi}
\end{eqnarray}
composed by fermion creation/annihilation operators in pseudo-spin space, $\beta^\dagger_{\kv n\spin}/\beta_{\kv n\spin}$. The $[\Gamma_l]_{\spin_1\spin_2}$ matrices in Eq.~(\ref{eq:Psi}) are constructed from the Pauli matrices $\sigma_l$ by
\begin{eqnarray}
 \Gamma_l&=&\frac{1}{\sqrt{2}}\sigma_l i \sigma_y .
\end{eqnarray}
This is analogous to the ${\bf d({\bf k})}$-vector ~\cite{SigristUeda} in the {\it pseudo-spin} space. 
In Eq.~(\ref{eq:Psi}), $s_0,s_y=-1$ and $s_x,s_y=+1$ and repeated indices are summed over. Only intraband Cooper pairing is included, as implied by the $\delta$-function in Eqs.~(\ref{eq:Psi}). The explicit form of the pairing kernel $\Gamma_{l,l'}(n,\kv;n',\kv')$ along with additional technical details can be found in the Supplementary Material of Ref.~\cite{PRL}.

\begin{figure*}[t]
 \centering
  \includegraphics[angle=0,height=0.18\linewidth]{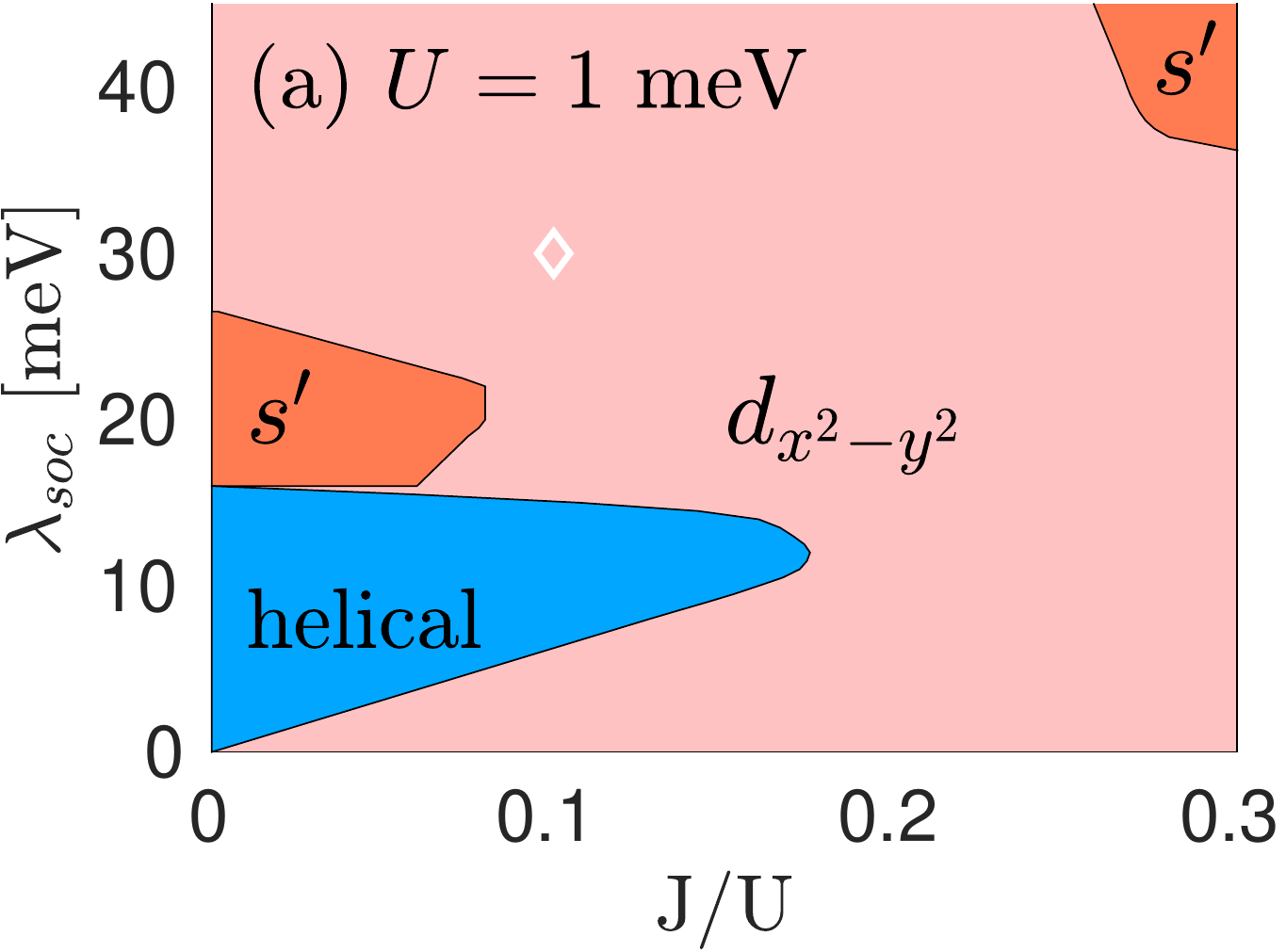}
   	\includegraphics[angle=0,height=0.18\linewidth]{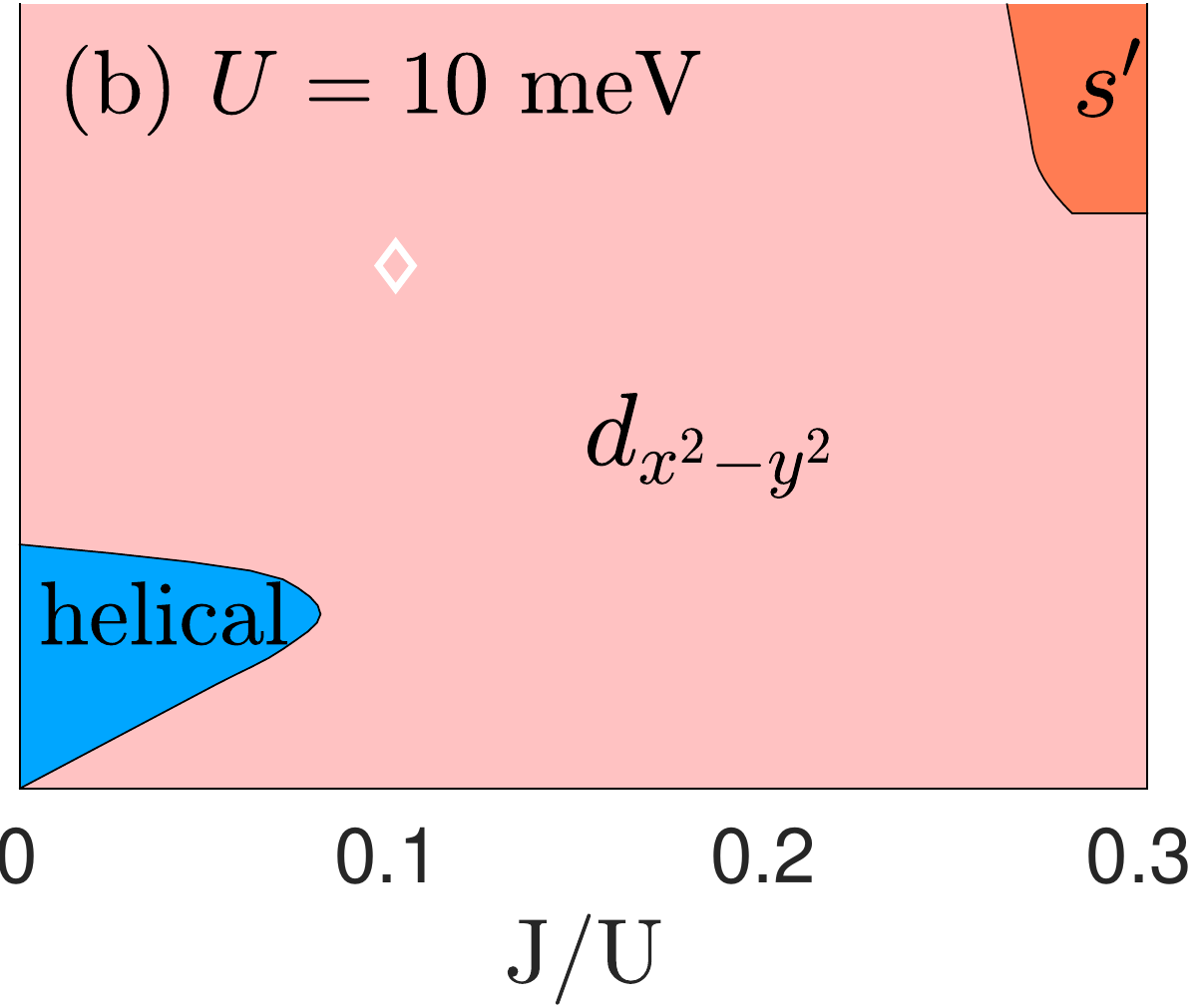}
   	\includegraphics[angle=0,height=0.18\linewidth]{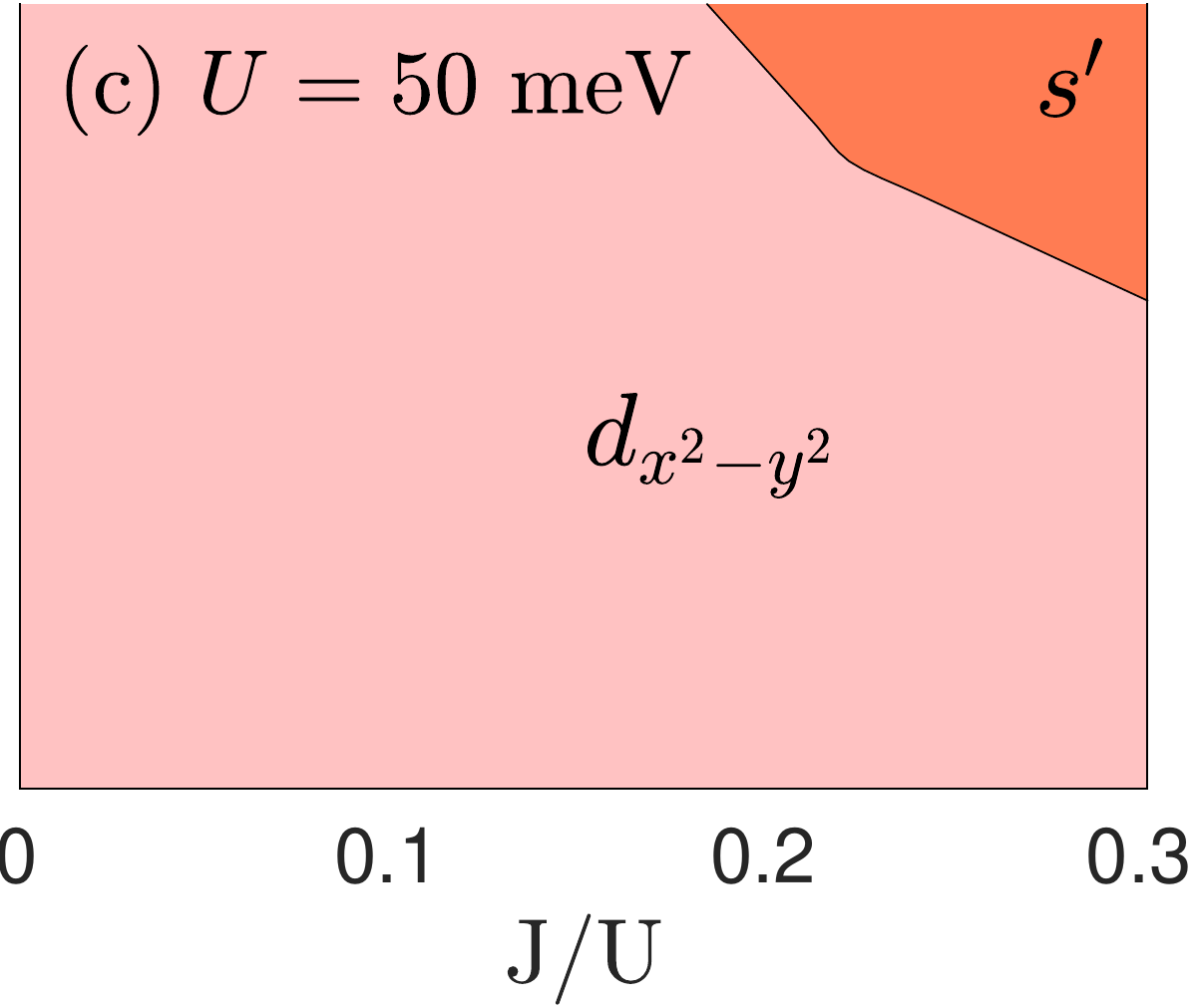}
   	   	\includegraphics[angle=0,height=0.18\linewidth]{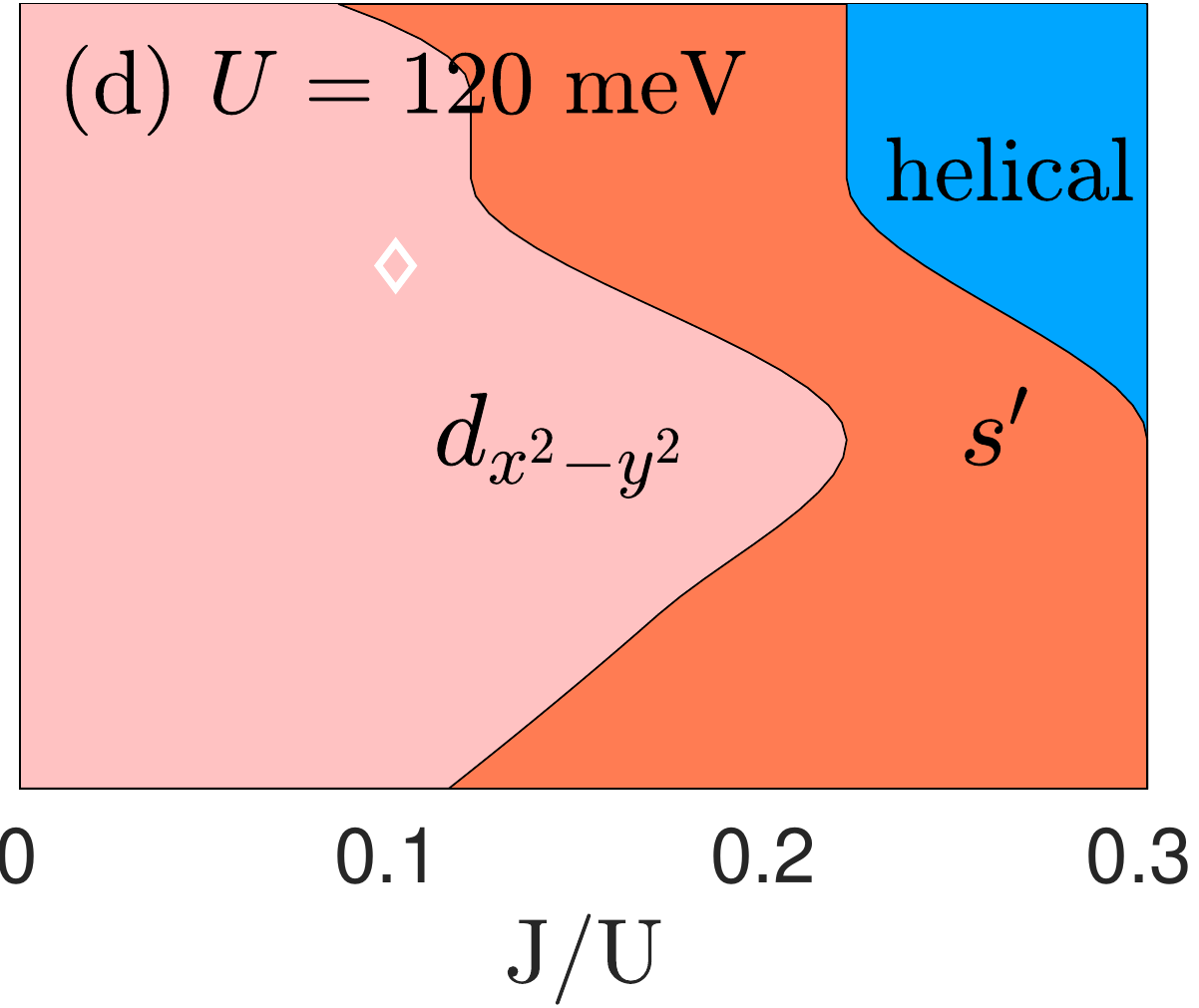}
\caption{Leading superconducting instability as a function of spin-orbit coupling $\lsoc$ and Hund's coupling $J$ for increasing value of $U$: $U=1,10,50,120$ meV with the inclusion of finite inter-orbital hybridization $g(\kv)=-4t^\prime\sin(k_x)\sin(k_y)$ with $t^\prime=4.4$ meV $(=0.05t_1)$ between the $xz$ and $yz$ orbital. There is overall agreement with the phase diagrams obtained for zero hybridization in Fig. 1, but small modifications are visible. White diamonds in (a,b,d) indicate positions for which we show the spectral $d_{x^2-y^2}$ gap in Fig.~\ref{fig:dwave}.}
\label{fig:LGE_Cobo_hyb}
 \end{figure*}

\begin{figure}[b]
 \centering
   	\includegraphics[angle=0,width=\linewidth]{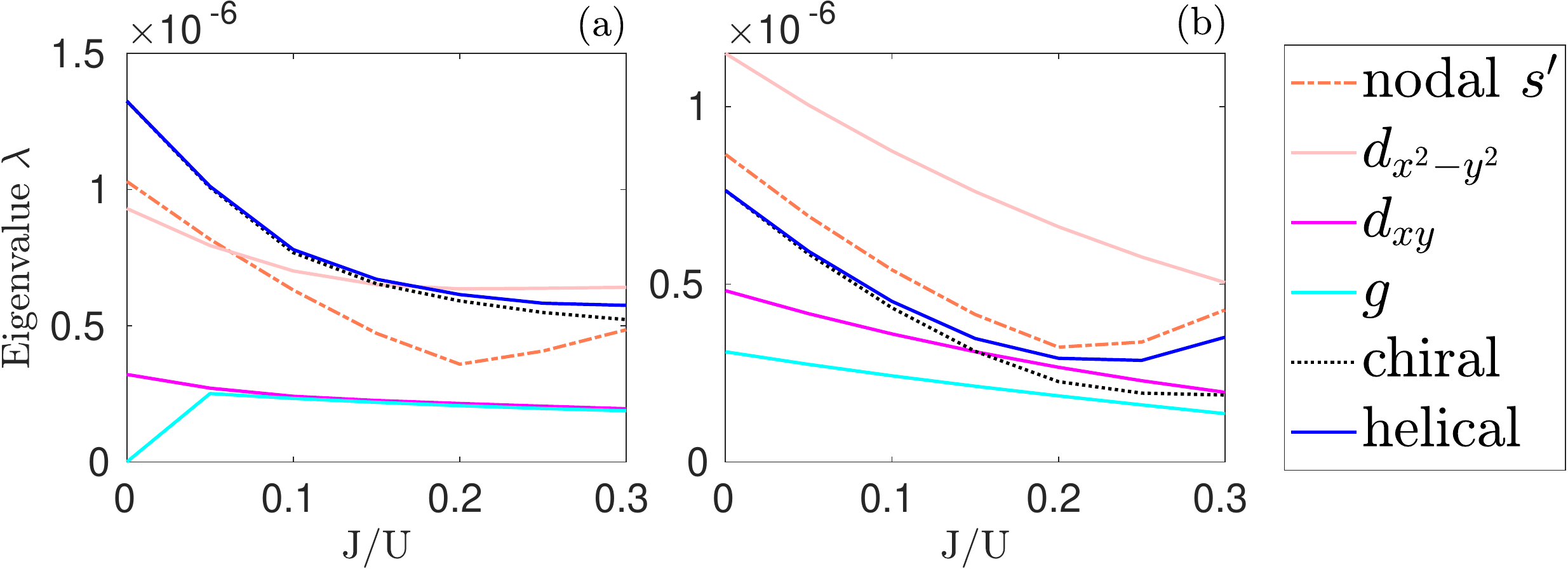}
\caption{Leading and subleading superconducting instabilities for $U=1$ meV with (a) $\lsoc=10$ meV and (b) $\lsoc=30$ meV (cuts are indicated by the dashed white lines in Fig. 1(a). Only the largest eigenvalue of each irreducible representation is depicted, i.e. higher order intermediate instabilities are not shown. }
\label{fig:subgaps}
 \end{figure}
 
\begin{figure}[b]
 \centering
   	\includegraphics[angle=0,width=\linewidth]{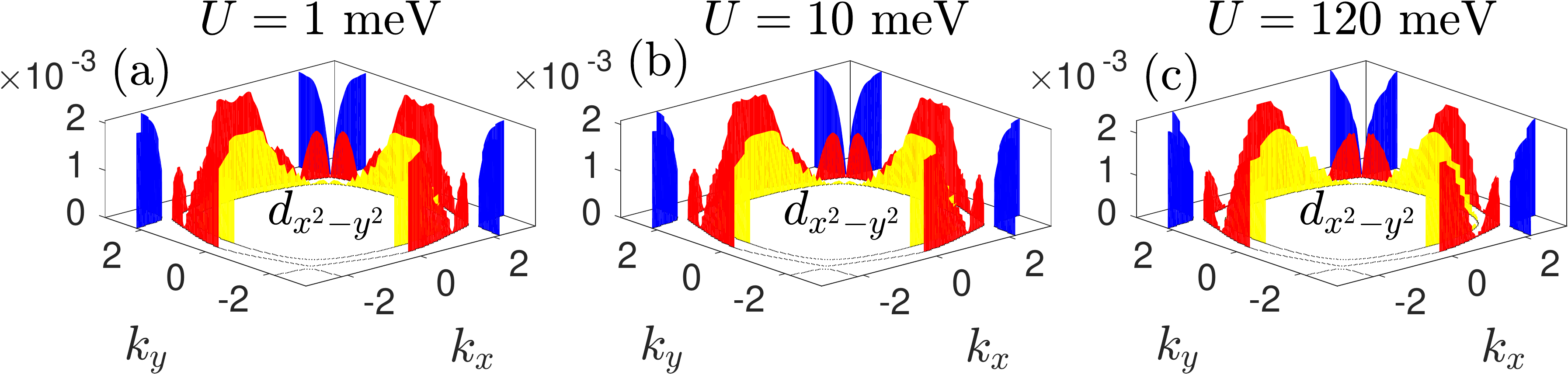}
\caption{Spectral gaps for the $d_{x^2-y^2}$ solutions for  $J/U=0.1$, $\lambda_{soc}=35$ meV and $t^\prime=4.4$ meV, indicated by white diamonds in Fig. 2 (a,b,d), from low-$U=1$ meV (a) through intermediate-$U=10$ meV  (b) to high-$U=120$ meV (c).
}
\label{fig:dwave}
 \end{figure}
\begin{figure}[t]
 \centering
   	\includegraphics[angle=0,width=\linewidth]{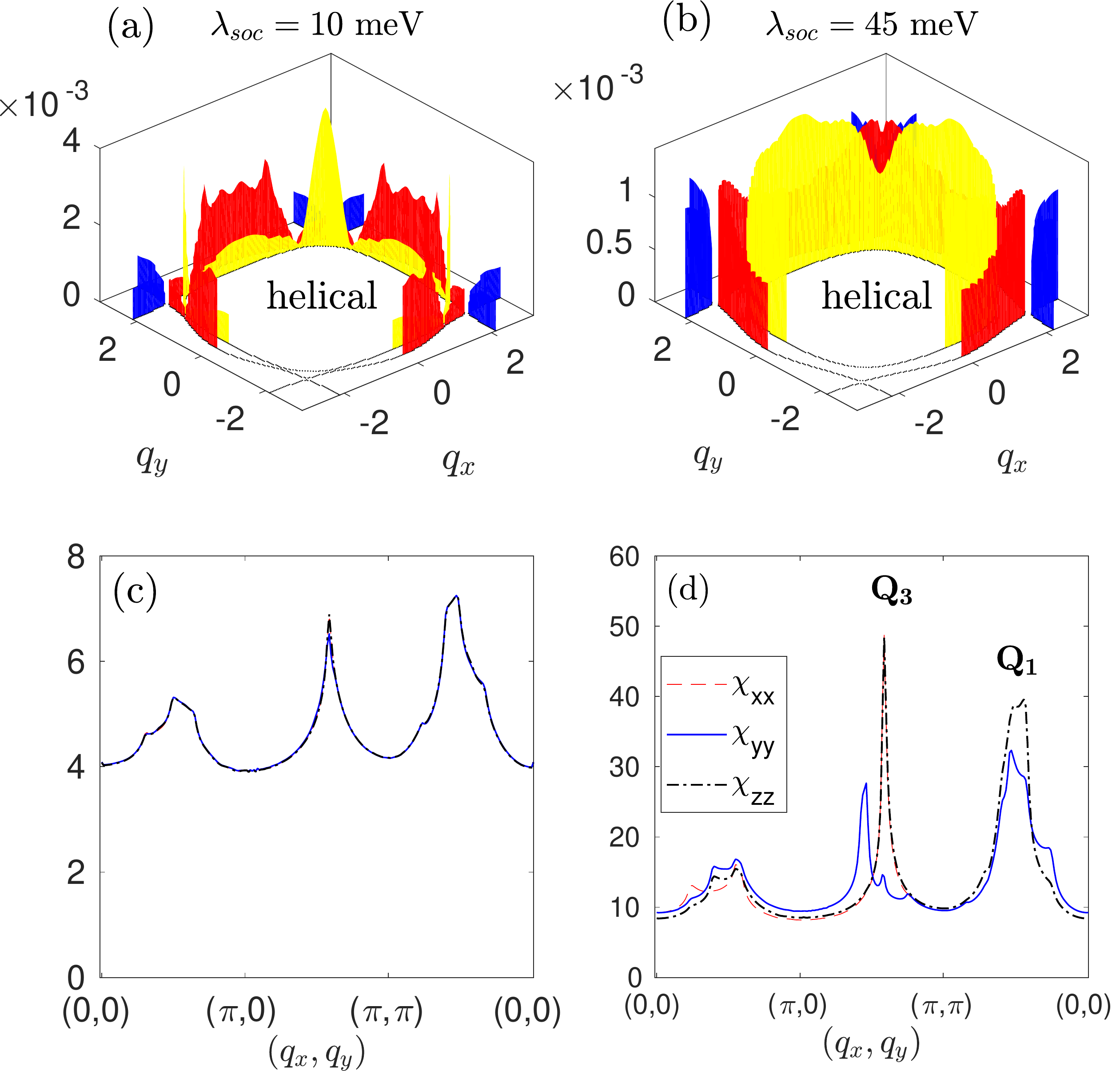}
\caption{Spectral gaps for the helical solutions in the  (a) low-$U$ regime ($U=1$ meV, $J/U=0.05$ and $\lambda_{soc}=10$ meV) and (b) high-$U$ regime ($U=120$ meV, $J/U=0.25$ and $\lambda_{soc}=35$ meV). (c,d) Physical spin susceptibilities for a cut through the Brillouin zone in the low- and high-$U$ regime, respectively.
}
\label{fig:helical}
 \end{figure}

The leading and sub-leading superconducting instabilities are determined from the linearized gap equation
\begin{eqnarray}
   -\!\int_{FS} \!\! d \kv_f^\prime \frac{1}{|v(\kv_f^\prime)|} \Gamma_{l,l'}(\kv_f,\kv_f^\prime)\Delta_{l'}(\kv_f^\prime)=\lambda \Delta_l(\kv_f),
  \label{eq:LGE}
\end{eqnarray}
where
%\begin{equation}
$    \Delta_l(n,\kv)=\frac{1}{2}\sum_{n^\prime, \kv^\prime,l'}\Gamma_{l,l'}(n,\kv;n',\kv^\prime)\langle \Psi_{l'}(n',\kv')\rangle. $
%\end{equation}
The integration in Eq.~(\ref{eq:LGE}) includes momenta at the Fermi surface of the three bands with $n$ uniquely defined by $\kv_f$, and $v(\kv_f)$ is the Fermi velocity at $\kv_f$. The structure of the leading superconducting instability is given by the eigenvector $\Delta_l(\kv_f)$ which corresponds to the largest eigenvalue $\lambda$.
As a result of tetragonal symmetry, we classify the even parity states $\Delta_0({\bf k})$ by 
$A_{1g} (s)$, $A_{2g} (g)$, $B_{1g} (d_{x^2-y^2})$ or $B_{2g} (d_{xy})$ symmetry. The odd parity states are either helical or chiral. There are four helical states obtained by superpositions of 
$\Delta_x({\bf k})$ and $\Delta_y({\bf k})$ which all have the pseudospin vector out of the plane. The chiral solution,
$\Delta_z({\bf k})$, is doubly degenerate and displays a pseudospin polarization in the plane.

In Fig.~\ref{fig:LGE_Cobo}, we show the leading instability displayed as a function of Hund's coupling and spin-orbit coupling, $(J/U,\lsoc)$, for increasing values of $U=1,10,50,120$ meV and zero hybridization, $t'=0$. 
For the smallest $U$, there is a regime of odd-parity helical superconductivity for the lowest values of $\lsoc$ and $J$. This is in agreement with the findings in Ref.~\onlinecite{WangKallin19}, which also report helical solutions in the low Hund's coupling regime.
As opposed to the reports in Ref.~\onlinecite{Zhang18}, we do not find a leading chiral solution, not even at the lowest values of $U=1$ meV. However, exactly at $J=t^\prime=0$, the helical and chiral solutions become degenerate as  expected from the analysis in Ref.~\onlinecite{WangKallin19}.

At large Hubbard-$U$, the picture changes completely. The helical solution at low SOC and $J$ disappears, but another helical solution becomes leading in the opposite limit of {\it large} SOC and $J$. The latter solution was reported recently in Ref.~\cite{PRL}.

The disappearance of the helical solution in the weak-coupling regime, Fig.~\ref{fig:LGE_Cobo}(a,b), is not easily understood from the underlying spin-fluctuation spectrum. As shown in Fig.~\ref{fig:LGE_Cobo}(e,f) the change in the spin fluctuations for $\lsoc=10$ meV and $J/U=0.05$ is almost indiscernible when increasing $U=1$ to $10$ meV, but nonetheless the change in leading superconducting instability is dramatic, shifting from odd-parity helical to even-parity $d_{x^2-y^2}$.
This shows that spin-fluctuation mediated pairing in \sruo ~is generally very  sensitive  not only to the details of the band structure as discussed in Ref.~\onlinecite{PRL}, but also the strength of the bare interaction. This fact was also highlighted in the recent paper by Zhang {\it et al.}~\cite{Zhang18}. In their paper, the low-$U$ regime shows a competition between helical and chiral pseudospin triplet while even-parity solutions are prominent at larger values of $U$. Note however, that the chiral solution is expected to become suppressed by orbital hybridization and Hund's coupling~\cite{WangKallin19}.

To investigate how the inclusion of a small hybridization between the $xz/yz$ orbitals changes the  phase diagram across the values of $U$, we show in Fig.~\ref{fig:LGE_Cobo_hyb} similar diagrams as in Fig.~\ref{fig:LGE_Cobo}, with the only modification of a small hybridization of $t^\prime=4.4$ meV. We see that while the effect is visible, in promoting the $s^\prime$ solution and also to some extend the helical solution over the $d_{x^2-y^2}$, the effect is very modest. 

In Fig.~\ref{fig:subgaps}, we show the spectrum of subleading instabilities as a function of Hund's coupling $J$ for two different values of SOC as indicated by dashed white lines in Fig.~\ref{fig:LGE_Cobo}(a) in the weak-coupling regime of $U=1$ meV. In general, the subleading instabilities are in close vicinity to the leading superconducting instability with the exception of the two even-parity solutions $d_{xy}$ and $g$-wave, which appear to be largely suppressed both in the weak- and strong-coupling regimes~\cite{PRL}. Overall, the most prominent solutions are found in the even-parity channel, especially the $d_{x^2-y^2}$-wave solution which dominates large regions of the phase diagrams in Fig. 1 and 2. The spectral gap of this solution is rather insensitive to the strength of $U$, which we show in Fig.~\ref{fig:dwave}. In all three cases of $U=1,10,120$ meV, the spectral $d_{x^2-y^2}$ remains roughly invariant, with symmetry-enforced diagonal nodes and strong gap suppression of the $\beta$-band gap in the neighborhood of the nodal directions, as visible from the yellow inner-most pocket in Fig.~\ref{fig:dwave}.  

Finally, we compare the low-$U$ and high-$U$ regime of odd-parity helical superconductivity in Fig.~\ref{fig:helical}. In the low-$U$ limit, the helical solution is found at relatively small SOC of $10$ meV and at small Hund's couplings. The magnetic susceptibility is almost spin-isotropic as a result of the small value of $\lsoc$ and $J$, as shown in Fig.~\ref{fig:helical}(c). The spectral gap in this regime resides mainly in the $xy$-orbital, i.e. the red $\gamma$-pocket as well as the large gap values along the zone diagonals on the yellow $\beta$-pocket in Fig.~\ref{fig:helical}(a). 
By contrast, the spectral gap of the helical state in the high-$U$ regime is much more evenly distributed on all three orbitals, as seen in Fig.~\ref{fig:helical}(b), which displays the spectral gap in the limit of large $U$, $\lsoc$ and $J$. Also, the susceptibility exhibits anisotropy between the in-plane and out-of-plane spin components at $\Qv_1\simeq(2\pi/3,2\pi/3)$ and $\Qv_3\simeq(\pi,2\pi/3)$. The spin anisotropy at $\Qv_1$ which increases with SOC and $J$~\cite{PRL}, is in agreement with neutron scattering reports~\cite{Braden04,Steffens19}. The peak structure at $\Qv_3$ is less reported in the literature, but ridge structures compatible with the peaks at $\Qv_3$ were reported from neutron scattering experiments in Ref.~\cite{Iida11}.

From the perspective of magnetic anisotropy of the spin fluctuations at $\Qv_1$, among the two proposals for helical gap structures, the solution in the large-SOC and large-coupling regime appears to be more in accordance with the experimental situation.  However, this solution does not exhibit nodal structure, which conflicts with experimental reports documenting (near-)nodal structure of the gap in \sruo~\cite{Hassinger17,Suzuki02,Ishida00,Bonalde00,Deguchi04,Suderow98}. In addition, the helical solutions do not conform with reports of time-reversal symmetry breaking~\cite{Luke1998,Kapitulnik09}.
We therefore  highlight  the fact that the even-parity solutions $d_{x^2-y^2}$ and nodal $s^\prime$-wave are prominent candidates for spin-fluctuation mediated superconductivity in \sruo ~which, like the helical solutions, also comply with the recent development in nuclear magnetic resonance results~\cite{Pustogow19,IshidaPreprint19}.
The near-degeneracy of these two solutions in large regions of the $(\lsoc,J/U)$ phase space, in particular for large values of $U$, lead to the proposal of a nodal, time-reversal broken, even-parity solution of the form $s^\prime+i d_{x^2-y^2}$.

In summary, we have provided additional numerical results for the leading superconducting instabilities in \sruo ~from the perspective of spin-fluctuation driven superconductivity. The modelling was restricted to two dimensions and included realistic band structures and spin-orbit coupling. We have explored the very weak-coupling regime, and discussed the gap structures of the different helical solutions appearing within this framework. In agreement with Ref. \cite{PRL}, the gap structure most compatible with the experimental situation appears to be even-parity solutions, possibly in the form of complex linear combinations due to accidental degeneracy.   

We acknowledge support from the Carlsberg Foundation. 
% remove following command to only show cited references
\nocite{*}
 \bibliography{bibliography_sr2ruo4}
\end{document}